\documentclass[aps,prl,twocolumn,superscriptaddress,footinbib]{revtex4-2}
\pdfoutput=1
\usepackage{amsmath}
\usepackage{graphicx}
\usepackage{amssymb}
\usepackage{enumitem}
\usepackage{mathrsfs}
\usepackage[linktocpage=true]{hyperref}

\newcommand{\ex}{\mathrm{e}}

\newcommand{\cycle}{\Upsilon}
\newcommand{\PP}{P}

\usepackage{color}

\def\nn{\nonumber}
\newcommand{\ii}{\mathrm{i}}

\begin{document}

\title{Entropy functions for accelerating black holes}

\author{Andrea Boido}
\affiliation{Mathematical Institute, University of Oxford, Woodstock Road, Oxford, OX2 6GG, U.K.}
\author{Jerome P. Gauntlett}
\affiliation{Blackett Laboratory, Imperial College, Prince Consort Road, London, SW7 2AZ, U.K.}
\author{Dario Martelli}
 \affiliation{Dipartimento di Matematica, 
 Universit\`a di Torino, 
Via Carlo Alberto 10, 10123 Torino, Italy}
\affiliation{INFN, Sezione di Torino,
 Via Pietro Giuria 1, 10125 Torino, Italy}
\author{James Sparks}
\affiliation{Mathematical Institute, University of Oxford, Woodstock Road, Oxford, OX2 6GG, U.K.}

\begin{abstract}
\noindent  We introduce an entropy function for supersymmetric accelerating black holes in 
$AdS_4$, that uplift on general Sasaki-Einstein manifolds $X_7$ to solutions of M-theory. 
This allows one to compute the black hole entropy without knowing the explicit solutions. 
A dual holographic microstate counting would follow from computing certain supersymmetric partition functions of Chern-Simons-matter theories compactified on a spindle. 
We make a general prediction for a class of such partition functions in terms of 
``blocks'', with each block being constructed from the partition function on a three-sphere.

\end{abstract}

\maketitle

\section{Introduction}\label{sec:intro}
Obtaining a precise microstate counting interpretation of black hole entropy is one of the major achievements of string theory. 
This was first studied in the context of supersymmetric and asymptotically flat black holes in
\cite{Strominger:1996sh} and this led to an enormous literature of further work. 
More recently, starting with \cite{Benini:2015eyy, Benini:2016rke}, there has been a growing body of similar work in the context of supersymmetric and
asymptotically $AdS$ black holes. In contrast to the earlier work, where it is the Cardy formula 
that underlies the microstate counting, instead the results for black holes in $AdS$ use holography and exact localization 
results for supersymmetric partition functions.

The present paper builds on \cite{Couzens:2018wnk, Gauntlett:2019roi, Hosseini:2019ddy, Kim:2019umc}, 
which introduced an entropy function for a large class of supersymmetric $AdS_4$ black holes 
 in M-theory. The seven-dimensional internal space $X_7$ is taken to be an arbitrary 
 Sasaki-Einstein manifold, with the three-dimensional holographic duals being 
 Chern-Simons-matter theories. The entropy function, similar in spirit to that 
 of Sen \cite{Sen:2005wa}, allows one to compute the 
 entropy without knowing the explicit supergravity solutions: the inputs 
 are only $X_7$, the topology of the black hole horizon $\Sigma$, taken to be a Riemann surface,
 and the magnetic  charges. This 
 entropy was shown to match a dual computation using the localization results of 
 \cite{Hosseini:2016tor, Hosseini:2016ume}, for infinite families of black holes.
 
In this note we extend \cite{Couzens:2018wnk, Gauntlett:2019roi, Hosseini:2019ddy} to  \emph{accelerating} black holes in $AdS_4$.
This leads to a number of novel features \cite{Ferrero:2020twa}: 
the black holes have different horizon topology, with conical deficit angles; 
supersymmetry is preserved  in a novel way; and 
when the deficit angles are appropriately quantized, 
so that the horizon $\Sigma$ is an orbifold known as a \emph{spindle}, remarkably
the uplifted $D=11$ solutions are completely smooth on and 
outside the horizon. 

We will 
explain how to compute the entropy 
function for a general class of accelerating $AdS_4$ black holes, 
which takes a simple ``gravitational block'' form, 
 vastly extending \cite{Hosseini:2019iad,Hosseini:2021fge,Faedo:2021nub}.  This leads to a 
general prediction for the 
partition functions of supersymmetric Chern-Simons-matter theories compactified on a spindle,  
with magnetic fluxes switched on 
for flavour and baryonic global symmetries,
in the large $N$ limit\footnote{It would be of much interest to generalise the entropy functions of this paper
to incorporate higher-derivative corrections and hence go beyond the large $N$ limit.}.
We also point out a striking relation of our entropy 
function to the on-shell action of the black holes in various examples, and comment 
on including angular momentum and electric charges in this formalism.

\section{Supersymmetric $AdS_2$ solutions}\label{sec:1}

Our starting point is the following general class of 
supersymmetric $AdS_2$ solutions to $D=11$ supergravity
introduced in \cite{Kim:2006qu} and clarified in \cite{Gauntlett:2007ts}:
\begin{align}\label{11Dsolution}
d s^2_{11} & = \ex^{-2B/3}\left(d s^2_{AdS_2} +d s^2_{Y_9}\right)\, ,\nn\\
G & = \mathrm{vol}_{AdS_2}\wedge \left[J-d(\ex^{-B}\eta)\right] \, .
\end{align}
Here $d s^2_{AdS_2}$ is a unit radius metric on $AdS_2$, with volume 
form $ \mathrm{vol}_{AdS_2}$ and  $G$ is the $D=11$ four-form. 
The GK space ${Y_9}$ has a canonically defined Killing vector field, $\xi$, 
called the \emph{R-symmetry vector}, and it plays a central role. 
The metric on $Y_9$ takes the form
\begin{align}\label{metricY}
d s^2_{Y_9} = \eta^2 + \ex^B  d s^2_T\, ,
\end{align}
where the one-form $\eta$ is dual to $\xi$, $ds^2_T$ is a K\"ahler metric transverse to $\eta$, with associated
K\"ahler two-form $J$ and Ricci two-form $d\eta=\rho$. 
The function $\ex^B = \frac{1}{2}R$, 
where $R>0$ is the scalar curvature of the K\"ahler metric.

The metric and four-form in \eqref{11Dsolution} give supersymmetric solutions to $D=11$ supergravity provided also
\begin{align}\label{PDE}
\Box R = \frac{1}{2}R^2 - R_{ab}R^{ab}\, ,
\end{align}
where $R_{ab}$ denotes the Ricci tensor for the K\"ahler metric, and $\Box$ is the Laplacian operator. 
However, to define our entropy function 
we wish to go \emph{off-shell} \cite{Couzens:2018wnk}, 
and in particular we will  not directly impose \eqref{PDE} in what follows. 

\section{Near horizon limits of black holes}\label{sec:2}

For the solutions of interest 
the $D=11$ vacuum solution, without the black hole,  
is $AdS_4\times X_7$, where $X_7$ is a Sasaki-Einstein manifold. 
A putative black hole may then carry conserved charges 
associated to various massless $U(1)$ gauge fields in $AdS_4$. The latter 
arise from Kaluza-Klein reduction on $X_7$,  either from isometries of $X_7$
(``flavour symmetries") or from homology
cycles (``baryonic symmetries").

Sasaki-Einstein manifolds $X_7$ have a $U(1)^s$ isometry, where necessarily 
$s\geq 1$, and we may choose an associated normalized basis 
of Killing vector fields $\partial_{\varphi_i}$, $i=1,\ldots,s$. 
$X_7$ is equipped with a 
Killing spinor, and without loss of generality we choose 
the basis so that this spinor has charge 
$\tfrac{1}{2}$ under $\partial_{\varphi_1}$, 
and is uncharged under the remaining vector fields. 
Via the Kaluza-Klein mechanism, massless $U(1)$ gauge fields $A_i$ in $AdS_4$ are obtained by
gauging $d\varphi_i\rightarrow d\varphi_i+A_i$ in the metric on $X_7$, together 
with adding a corresponding term to the $D=11$ six-form potential $C_6$, given in \eqref{deltaC6} below,
where $dC_6=*_{11}G$.
On the other hand, if $\Sigma_I\subset X_7$ 
form a basis of five-cycles, $I=1,\ldots,b_5(X_7)= \dim H_5(X_7,\mathbb{R})$, then 
 reducing $C_6$ on 
each five-cycle $\Sigma_I$ also leads to massless $U(1)$
gauge fields $A_I$ in $AdS_4$. Altogether we have the linear perturbation
\begin{align}\label{deltaC6}
\delta C_6 = \sum_{i=1}^s A_i\wedge \omega_i + \sum_{I=1}^{b_5(X_7)}A_I\wedge \omega_I\, .
\end{align}
Here both $\omega_i$ and $\omega_I$ are co-closed five-forms on $X_7$, but 
$\omega_I$ is closed while $d\omega_i= \partial_{\varphi_i}\lrcorner\,  \mathrm{vol}_{X_7}$ \cite{Benvenuti:2006xg}, for a 
suitably normalized volume form on $X_7$. Notice that in \eqref{deltaC6} we are  free to 
shift $\omega_i\rightarrow \omega_i + \sum_{I} c_i^I \omega_I$, 
for arbitrary constants $c_i^I$, which is precisely the freedom to 
mix baryonic symmetries into flavour symmetries in field theory. 
This correspondingly shifts $A_I \rightarrow A_I - \sum_i c_i^I A_i$ and hence the notion of baryonic fluxes in
 the reduced theory on $AdS_4$. 
 
Consider introducing a supersymmetric extremal black hole into this 
$AdS_4$ vacuum. The near horizon limit should be  $AdS_2\times \Sigma$, 
where the two-dimensional surface $\Sigma$ is the black hole horizon \footnote{While this is generically expected to be true, it is important to understand the necessary and sufficient conditions.}. 
For an accelerating black hole we take $\Sigma$ to be a \emph{spindle}  \cite{Ferrero:2020twa}. This is topologically a two-sphere, 
but with conical deficit angles $2\pi(1-1/m_\pm)$ at the poles, 
specified by two coprime positive integers $m_\pm$.  The non-accelerating case 
is  recovered simply by setting $m_\pm =1$, so $\Sigma=S^2$.

Now consider turning on \emph{flavour} magnetic charges, for the 
gauge fields 
originating from isometries of $X_7$, with
\begin{align}\label{flavourflux}
\frac{1}{2\pi}\int_\Sigma dA_i  = \frac{p_i}{m_-m_+}\, ,
\end{align}
the magnetic flux through the horizon. 
This precisely fibres $X_7$ over $\Sigma$ to give a GK geometry of the form
\begin{align}\label{fibration}
X_7\hookrightarrow Y_9 \rightarrow \Sigma\, .
\end{align}
The fibration is well-defined \footnote{$Y_9$ is free of orbifold singularities provided the $p_i$ are coprime to both of $m_\pm$.} 
when the flavour magnetic charges $p_i$ are integers \cite{Ferrero:2021etw}. 
Imposing supersymmetry requires \cite{Ferrero:2021etw} that 
\begin{align}\label{p1}
p_1 = -\sigma m_+-m_-\, ,
\end{align}
where recall that the first copy of $U(1)$ is singled out by the Killing spinor being 
charged under this symmetry. Here $\sigma=\pm1$ are called 
\emph{twist} and 
\emph{anti-twist}, respectively. 

On the other hand, 
the $D=11$ seven-form flux 
satisfies the Dirac quantization condition
\begin{align}\label{fluxDirac}
\frac{1}{(2\pi \ell_p)^6}\int_{\cycle} dC_6  = N_\cycle\in \mathbb{Z}\, ,
\end{align}
where $\ell_p$ is the $D=11$ Planck length, and $\cycle\subset Y_9$ is 
any seven-cycle. When $Y_9$ takes the fibred form \eqref{fibration}
 there is a distinguished such cycle, namely a copy $\cycle=X_7$ of the fibre, 
 and we identify $N\equiv N_{X_7}$ with the number of M2-branes 
generating the original $AdS_4\times X_7$ vacuum \footnote{We must have $N=m_+N^{X_+}=m_-N^{X_-}$ with $N^{X_\pm}\in\mathbb{Z}$ and hence $N=m_+ m_-{\mathcal N}_0$ with ${\mathcal N}_0\in\mathbb{Z}$  \cite{Boido:2022mbe}.}. If we pick representatives of the 
five-cycles $\Sigma_I\subset X_7$ that are invariant under the $U(1)^s$ action, 
then via \eqref{fibration} these will fibre over the spindle $\Sigma$ to give 
associated seven-cycles $\cycle_I\subset Y_9$. We denote the corresponding flux numbers in \eqref{fluxDirac} 
as $N_I\equiv N_{\cycle_I}$, and analogously to \eqref{flavourflux} define \emph{flux}
magnetic charges $\PP_I\equiv N_I/N$ \footnote{The normalization factor of $N$ is due to the 
fact that baryonic operators, dual to M5-branes wrapped on $\Sigma_I$, arise as $N\times N$ determinants in Chern-Simons-matter duals, 
and $P_I$ is then the charge of the associated field.}. 
Notice that via \eqref{deltaC6} these fluxes will in general include contributions 
from the flavour magnetic charges $p_i$ in \eqref{flavourflux}, and 
also \emph{baryonic} magnetic charges $\frac{1}{2\pi}\int_\Sigma dA_I$. 
However, as explained above, defining the latter in general requires (arbitrary) 
choices, and so we instead parametrize the baryonic 
magnetic charges of the black hole via the $P_I$. 
This accounts for all quantized fluxes on $Y_9$.

\section{Entropy function}\label{sec:3}

We have seen how fixing the magnetic charges 
$p_i$, $\PP_I$ of the black hole 
encodes 
the twisting of the fibration \eqref{fibration}, and also 
quantized flux numbers \eqref{fluxDirac} for the corresponding near horizon 
$AdS_2\times Y_9$ solution. This can be related to 
geometric data on 
$Y_9$ as follows  \cite{Couzens:2018wnk}. First, evaluating the left hand side 
of \eqref{fluxDirac} on the background \eqref{11Dsolution} gives
\begin{align}\label{fluxquantize}
 \frac{1}{(2\pi\ell_p)^6}\int_\cycle  \eta \wedge \rho \wedge \frac{1}{2}J^2 = N_\cycle\, ,
\end{align}
while imposing that the \emph{integral} of \eqref{PDE} over $Y_9$ holds 
gives
\begin{align}\label{constraint}
\int_{Y_9}\eta\wedge \rho^2 \wedge J^2 = 0\, .
\end{align}

The integrals \eqref{fluxquantize}, \eqref{constraint} are functions of  K\"ahler class parameters $[J]$ and the R-symmetry vector, which we may write as
\begin{align}\label{Rbasis}
\xi = \sum_{\mu = 0}^s b_\mu \partial_{\varphi_\mu}\, .
\end{align}
Here $\partial_{\varphi_0}$ is a Killing vector field rotating the spindle $\Sigma$, fixing its poles. 
The K\"ahler class parameters
lie in the basic cohomology 
$[J]\in H^2_B(\mathcal{F}_\xi)$ associated to the foliation $\mathcal{F}_\xi$ defined by 
$\xi$. 
One can show that the total number of such parameters is $\dim H_5(X_7,\mathbb{R}) + 2$. 
On the other hand, fixing the flux magnetic charges $\PP_I$, together with $N$ and imposing  \eqref{constraint}, 
imposes the same number of constraints. Although we do not have a general argument, in all examples 
fixing $p_i$, which determines the topology of $Y_9$, 
together with $\PP_I$ and $N$ fixes all the K\"ahler class parameters. 
This leaves the R-symmetry vector \eqref{Rbasis} still unspecified, 
apart 
from the constraint $b_1=1$ which corresponds to the Killing spinor necessarily having charge~$\tfrac{1}{2}$.

The main result of \cite{Couzens:2018wnk} is that 
solutions to the PDE 
\eqref{PDE} extremize the \emph{entropy function}
\begin{align}\label{Sdef}
\mathscr{S} \equiv \frac{4\pi }{(2\pi)^8 \ell_p^9}\int_{Y_9}\eta\wedge \rho\wedge \frac{1}{3!}J^3\, .
\end{align}
For fixed $X_7$, spindle data $m_\pm$, magnetic charges $p_i$, $\PP_I$ and $N$, 
 we have
$\mathscr{S}=\mathscr{S}(b_\mu)$ is a function only
of the R-symmetry vector. The near horizon 
$AdS_2$ solution necessarily 
extremizes this, as a function of $(b_0,b_1=1,b_2,\ldots,b_s)$, 
with the black hole entropy  $S_{\mathrm{BH}}=\mathscr{S}(b_\mu^*)$ 
being the entropy function evaluated at the critical point.

\section{Gravitational blocks}\label{sec:4}

Using Stokes' theorem one can show that the entropy function \eqref{Sdef} 
can be written in the ``block'' form
\begin{align}\label{Spreblock}
\mathscr{S} = \frac{4\pi }{(2\pi)^7 \ell_p^9}\frac{b_1}{b_0}\left[\mathrm{Vol}(X_7^+)-\mathrm{Vol}(X_7^-)\right]\, ,
\end{align}
(see \cite{Boido:2022mbe} for details).
Here $X_7^\pm$ are the copies of $X_7$ over the two poles of the spindle \footnote{More precisely 
these fibres are $X_7/\mathbb{Z}_{m_\pm}$, so $X_7^\pm$ are really covering spaces of the fibres. 
The orientations of $X_7^\pm$ are discussed in more detail in \cite{Boido:2022mbe}.}, 
and $\mathrm{Vol}(X_7^\pm)=\int_{X_7^\pm}\eta\wedge \frac{1}{3!}J^3$ is the volume induced 
by the choice of K\"ahler class. 
For toric $X_7$, which by definition have $s=4$ and so at least $U(1)^4$ isometry, 
this was called the ``master volume'' in \cite{Gauntlett:2019roi} and
references
 \cite{Gauntlett:2019roi, Boido:2022mbe} describe in detail how to compute this master volume in terms of toric data. 
 
In practice \eqref{fluxquantize}, \eqref{constraint} are quadratic in the K\"ahler class parameters, and for more than one such parameter 
it is typically difficult to solve for $[J]$ in closed form, and thus obtain the entropy function $\mathscr{S}$ 
as described after equation \eqref{Sdef}. In the remainder of this paper \footnote{Tools for analysing toric examples 
are developed in \cite{Boido:2022mbe}} we will hence focus on a
restricted, but still very rich, class of examples that we refer to as \emph{flavour twists}. 
This generalizes a similar class studied in \cite{Hosseini:2019ddy}, where by definition 
we impose that $[J|_{X_7^{\pm}}]\propto [\rho |_{X_7^{\pm}}]$.
It can be shown 
that \eqref{Spreblock} leads to the result
 \begin{align}\label{Sflavour}
 \mathscr{S}  = \frac{8\pi^3 N^{3/2}}{3\sqrt{6}b_0}\Bigg(\frac{1}{\sqrt{\mathrm{Vol}_S(X_7)\mid_{\vec{b}^+}}}
 -\frac{\sigma}{\sqrt{\mathrm{Vol}_S(X_7)\mid_{\vec{b}^-}}}\Bigg)\, .
 \end{align}
 Here $\sigma=\pm1$ as in \eqref{p1}, and $\mathrm{Vol}_S(X_7)$ is the Sasakian volume of 
 $X_7$, introduced in \cite{Martelli:2006yb}. This is a function only of the 
 R-symmetry vector $\vec{b}=(b_1=1,b_2,\ldots,b_s)$ ({\it i.e.} excluding the $b_0$ spindle direction), 
 with
 \begin{align}
 \vec{b}^+ & \equiv \vec{b} - \frac{b_0}{m_+}(1,-a_+p_2,\ldots,-a_+ p_s)\, ,\nonumber\\
  \vec{b}^- & \equiv \vec{b} + \frac{b_0}{m_-}(\sigma ,-a_-p_2,\ldots,-a_- p_s)\, ,
 \end{align}
where $a_\pm$ are integers satisfying 
$a_- m_+ + a_+ m_- = 1$. 
Such $a_\pm$ exist by Bezout's lemma, as $m_\pm$ are coprime. 
They are not unique, but different choices amount to a different choice of 
basis for the $U(1)^{s+1}$ action on $Y_9$, with generators $\partial_{\varphi_\mu}$, 
$\mu=0,1,\ldots,s$, and the black hole entropy 
which extremizes \eqref{Sflavour} is independent of this choice. 

For fixed $X_7$, the entropy function \eqref{Sflavour} 
is manifestly a function of only $m_\pm$, $N$ the flavour magnetic charges $p_i$, 
and the R-symmetry vector $(b_0,b_1=1,b_2,\ldots,b_s)$. The flavour 
charges $p_2,\ldots,p_s$ are here arbitrary, but in this class of examples 
the flux charges $\PP_I$ are determined by the remaining data. Specifically, one can show \cite{Boido:2022mbe}
\begin{align}\label{PI}
\PP_I = \frac{\pi}{3b_0}\left(\left.\frac{\mathrm{Vol}_S(\Sigma_I)}{\mathrm{Vol}_S(X_7)}\right|_{\vec{b}^+}-\left.\frac{\mathrm{Vol}_S(\Sigma_I)}{\mathrm{Vol}_S(X_7)}\right|_{\vec{b}^-}\right)\, ,
\end{align}
where these are again Sasakian volumes. Various methods for computing these volumes, for 
different classes of $X_7$, were given in \cite{Martelli:2006yb}, including using toric geometry, 
a fixed point theorem and a limit of an equivariant index.

The entropy function \eqref{Sflavour} may thus be written down 
for infinite families of accelerating $AdS_4$ black holes in M-theory, 
with general flavour magnetic charges $p_i$, and extremized 
over the R-symmetry vector to obtain the entropy. We 
present some examples in the next section. 

The $AdS_4/CFT_3$ correspondence relates the free energy of the dual field theory (here typically Chern-Simons-matter theories) 
	on the three-sphere $S^3$ \cite{Martelli:2011qj, Cheon:2011vi,Jafferis:2011zi} to a gravitational quantity via $\mathcal{F}_{S^3}(\vec{b})=2^{1/2}3^{-3/2}\pi^3/\sqrt{\mathrm{Vol}_S(X_7)\mid_{\vec{b}}}$. This has been shown in many examples, although we are not aware of a general proof.
We may then write \eqref{Sflavour} as
\begin{align}\label{Sfree}
\mathscr{S}=\frac{4}{b_0}\left[\mathcal{F}_{S^3}(\vec{b}^+) - \sigma\mathcal{F}_{S^3}(\vec{b}^-) \right]\, .
\end{align}
Here the free energy blocks 
are functions of the shifted R-symmetry vectors $\vec{b}^\pm$, which in 
field theory 
correspond
to certain shifted trial R-symmetry assignments for the fields. 

Finally, consider setting $m_\pm=1$, so that the horizon $\Sigma=S^2$ and there is no acceleration, 
and also taking the limit $b_0\rightarrow 0$ so that the R-symmetry vector is purely tangent to $X_7$. From 
\eqref{Sflavour} (or \eqref{Sfree}) we then obtain
\begin{align}\label{Sround}
\mathscr{S} = {4}\sum_{i=1}^s p_i \frac{\partial}{\partial b_i}\sqrt{\frac{2\pi^6}{27\mathrm{Vol}_S(X_7)\mid_{\vec{b}}}}N^{3/2}\, ,
\end{align}
and here we should take $b_1=1$ after taking the derivative. 
This recovers the results of \cite{Gauntlett:2019roi, Hosseini:2019ddy}, where 
the derivative operator precisely acts on $\mathcal{F}_{S^3}(\vec{b})$. 

\section{Examples}

The are two particularly interesting 
classes of examples of the flavour twist construction described above, where in particular cases 
we may also make contact with various explicit solutions. The first is when there are no baryonic 
symmetries, {\it i.e.} $H_5(X_7,\mathbb{R})=0$. In this case $J$ is necessarily 
exact on $X_7$, and the condition \eqref{PI} is vacuous. A simple example 
is $X_7=S^7$, for which $s=4$ and in a natural choice of basis for $U(1)^4$ we have 
\begin{align}
\left.\mathrm{Vol}_S(S^7)\right|_{\vec{b}}\,  = \frac{\pi^4}{3b_2b_3b_4(b_1-b_2-b_3-b_4)}\, .
\end{align}
One can check that the entropy function \eqref{Sflavour} 
agrees with the conjectured entropy function in \cite{Faedo:2021nub} (after a simple linear change of variable), and moreover 
extremizing to obtain the entropy the result agrees with the explicit near horizon supergravity solutions in \cite{Couzens:2021cpk}. 
Instead the non-accelerating result \eqref{Sround} was in this case already known  \cite{Hosseini:2019use} 
to reproduce the entropy 
of the family of STU supergravity black hole solutions in \cite{Benini:2016rke}. 
Another example in this class, treated in \cite{Boido:2022mbe}, is $X_7=V_{5,2}$, for which no explicit supergravity solutions are known. 

The second class of examples 
 are referred to as the \emph{universal anti-twist}. 
These  correspond to the 
explicit accelerating black hole solutions constructed in \cite{Ferrero:2020twa}. They are universal in the sense that 
the solutions  exist for 
arbitrary choice of Sasaki-Einstein $X_7$ with rational R-symmetry  vector (see \eqref{universalp} below). 
Moreover the solutions exist only in the anti-twist case, with $\sigma=-1$, as we shall see 
momentarily \footnote{Switching off rotation and electric charge gives solutions where the   spindle degenerates at the $AdS_4$ boundary  \cite{Ferrero:2020twa}.}. 
The universal anti-twist may be characterized geometrically 
by saying that the flavour twisting is only along the 
R-symmetry direction of the Sasaki-Einstein metric. This is equivalent to  imposing
\begin{align}\label{universalp}
\vec{p}\,  = \frac{p_1}{b_1^+}\, \vec{b}^+ = \frac{p_1}{b_1^-}\, \vec{b}^-\, .
\end{align}
Using a homogeneity property of the Sasakian volume, one can show \cite{Boido:2022mbe} that 
\eqref{Sflavour} leads to the simple result
\begin{align}\label{Suniversal}
\mathscr{S} = \frac{1}{4b_0}\left[(b_1^+)^2-\sigma (b_1^-)^2\right]\mathcal{F}_{S^3}\, .
\end{align}
Here  the free energy $\mathcal{F}_{S^3}=\mathcal{F}_{S^3}(\vec{b}^*)$ is computed using the (extremal) Sasaki-Einstein metric. 
In \eqref{Suniversal}
 one should set $b_1=1$ and extremize over $b_0$ to obtain the entropy. If $\sigma=+1$ there 
 are no extrema, forcing $\sigma=-1$. 
 One can check that the (positive) extremal value $S_{\mathrm{BH}}=\mathscr{S}^*$ is given by
 \begin{align}\label{Suniversalos}
S_{\mathrm{BH}} = \frac{(2m_-^2+2m_+^2)^{1/2}-m_--m_+}{2m_- m_+}\mathcal{F}_{S^3}\, ,
\end{align}
which precisely agrees with the 
 entropy of the explicit family of supersymmetric accelerating black holes in \cite{Ferrero:2020twa}. 
 
\section{On-shell action}

One might wonder how the entropy function we have introduced is 
related to other  approaches to computing 
black hole entropy, and the associated thermodynamics. 
An immediate issue for extremal black holes in $AdS_4$ is that 
the infinite $AdS_2$ throat leads to a (IR) divergence 
in the holographically renormalized on-shell action $I$, 
which is thus ill-defined without some form of regularization. 

In  \cite{Cassani:2021dwa} a \emph{complex} locus of supersymmetric 
but non-extremal accelerating black holes was considered, 
that in addition have non-zero rotation and electric charge.
This complex locus has
well-defined but 
complex action:
\begin{align}\label{Iuniversal}
I = \pm \frac{1}{\ii \pi} \left[\frac{\varphi^2}{\omega} + \left(\frac{m_--m_+}{4m_-m_+}\right)^2 \omega\right]\mathcal{F}_{S^3}\, .
\end{align}
Here the two signs correspond to two different complex branches, and $\varphi$, $\omega$ are chemical 
potentials associated to electric charge and rotation. These satisfy the constraint 
$\varphi=\frac{\chi}{4}\omega \pm \ii \pi$, where $\chi=(m_++m_-)/m_-m_+$ is the 
orbifold Euler characteristic of the spindle horizon. 
The entropy is obtained in a standard way from this, as minus the Legendre transform of $I$, 
passing from grand canonical to microcanonical ensemble. Remarkably 
\eqref{Suniversal} and \eqref{Iuniversal} satisfy $I=-\mathscr{S}$, via the change of variable
\begin{align}\label{changevariable}
\omega = \mp 2\pi \ii\, b_0\, .
\end{align}
The Legendre transform of $I$ thus extremizes 
$\mathscr{S}$, and since $\omega$ is a chemical potential 
for rotation of the horizon, and $b_0$ is the component of the R-symmetry vector rotating the horizon, \eqref{changevariable} is 
a  
natural identification.

On the other hand, $I$ is an on-shell quantity for the $AdS_4$ black holes, 
while $\mathscr{S}$ is an off-shell quantity for the associated near horizon 
$AdS_2$ solutions. It is therefore hard to see how these might be 
related physically, although by construction both are ``entropy functions'', in the sense 
that extremizing both gives the (same) black hole entropy. 
There is a similar relation between the entropy function \eqref{Sround}
in the case of $X_7=S^7$ and the on-shell action of the STU black holes computed in 
\cite{Cassani:2019mms} (see also \cite{Ferrero:2021ovq, Faedo:2021nub}), suggesting 
this relation is not accidental. 

\section{Angular momentum and electric charge}

It should be possible to generalize the analysis of this paper and also turn on both angular momentum $J$ 
and electric charges $q_i$, $Q_I$ for the $AdS_4$ black holes.
The fact that these are zero here is simply due to  
 \eqref{11Dsolution}: adding 
 rotation and electric charge modifies this ansatz \cite{Couzens:2020jgx}. 
 
 In \cite{Boido:2022mbe} some additional observables in gravity are also introduced, namely the geometric 
 R-charges \footnote{\cite{Boido:2022mbe} also shows that associated with $U(1)^s$ invariant
 non-trivial five-cycles $\Sigma_I$ on $X_7$, (16) can be recast in the form 
$\PP_I = \frac{1}{2b_0}\left(R_I^+-R_I^-\right)$}
 \begin{align}\label{Rpm}
 R_a^+ & \equiv \frac{4\pi m_+}{(2\pi \ell_p)^6 N}\int_{S^+_a} \eta\wedge \frac{J^2}{2!}\, , \nonumber\\ 
  R_a^- & \equiv \frac{4\pi \sigma m_-}{(2\pi \ell_p)^6 N}\int_{S^-_a} \eta\wedge \frac{J^2}{2!}\, .
 \end{align}
 Here $S_a^\pm$ are a set of $U(1)^s$-invariant supersymmetric five-submanifolds of the fibres $X_7^\pm$,
and note that these exist even when $\dim H_5(X_7,\mathbb{R})=0$. These geometric R-charges are dual to R-charges of baryonic operators associated with M5-branes wrapping the submanifolds. 
When   $X_7$ is toric, the cones over these are precisely the toric divisors in the Calabi-Yau cone $C(X_7)$, labelled by $a=1,\ldots,d$. 
In this toric case we have the identity \cite{Boido:2022mbe}
\begin{align}
\frac{1}{2}\sum_{a=1}^d \left(R_a^++ R_a^-\right) = 2 - \frac{m_--\sigma m_+}{m_+m_-}b_0\, .
\end{align}
In the universal anti-twist case, with the identification \eqref{changevariable}, 
we then note we may identify the chemical potential $\varphi=\pm \frac{\ii \pi}{4}\sum_{a=1}^4 (R_a^++R_a^-)$.

 For the special case of $X_7=S^7$, where the index $a$ may be identified with the flavour index $i$, we can define the 
\emph{master entropy function}
\begin{align}\label{Smodified}
S \equiv \mathscr{S} - \ii \left[4 b_0 J -\frac{1}{4} \sum_{a=1}^d (R_a^++ R_a^-) q_a\right]\mathcal{F}_{S^3}\, .
\end{align} 
Here $\mathscr{S}$ is the entropy function already introduced, depending on spindle data, magnetic charges 
and R-symmetry vector $b_0$, $b_i$, $i=1,\ldots,4$. In \eqref{Smodified} we have further
introduced angular momentum $J$, conjugate to $b_0$, and electric charges $q_a$, conjugate to the 
R-charges \eqref{Rpm}.  This is a natural generalization of the entropy function 
conjectured for the non-accelerating STU black holes in \cite{Benini:2016rke}, 
and moreover we have checked that extremizing $S$ and imposing that
 $S$ and the conserved charges 
are real, precisely leads to the entropy of the family of near horizon 
solutions constructed in \cite{Ferrero:2021ovq}. These were conjectured 
to be the near horizon limits of general
 dyonically charged, rotating and accelerating 
black holes in $AdS_4$ in STU gauged supergravity, which uplift on $X_7=S^7$ to solutions of M-theory. 
In the case with only a single dyonic pair of charges 
turned on for the graviphoton, these are precisely the black hole solutions in \cite{Cassani:2021dwa}.

More generally the flavour and baryonic charges 
are naturally combined in toric geometry via the index $a=1,\ldots,d$, and it is natural to 
conjecture that \eqref{Smodified} is the correct entropy function with general charges, not just for $X_7=S^7$ but for more general
classes of $X_7$.

\section*{Acknowledgments}
We thank Seyed Morteza Hosseini for helpful discussions.
This work was supported in part by STFC grants  ST/T000791/1 and 
ST/T000864/1.
JPG is supported as a Visiting Fellow at the Perimeter Institute. 

\bibliographystyle{apsrev4-1}


\begin{thebibliography}{38}%
\makeatletter
\providecommand \@ifxundefined [1]{%
 \@ifx{#1\undefined}
}%
\providecommand \@ifnum [1]{%
 \ifnum #1\expandafter \@firstoftwo
 \else \expandafter \@secondoftwo
 \fi
}%
\providecommand \@ifx [1]{%
 \ifx #1\expandafter \@firstoftwo
 \else \expandafter \@secondoftwo
 \fi
}%
\providecommand \natexlab [1]{#1}%
\providecommand \enquote  [1]{``#1''}%
\providecommand \bibnamefont  [1]{#1}%
\providecommand \bibfnamefont [1]{#1}%
\providecommand \citenamefont [1]{#1}%
\providecommand \href@noop [0]{\@secondoftwo}%
\providecommand \href [0]{\begingroup \@sanitize@url \@href}%
\providecommand \@href[1]{\@@startlink{#1}\@@href}%
\providecommand \@@href[1]{\endgroup#1\@@endlink}%
\providecommand \@sanitize@url [0]{\catcode `\\12\catcode `\$12\catcode
  `\&12\catcode `\#12\catcode `\^12\catcode `\_12\catcode `\%12\relax}%
\providecommand \@@startlink[1]{}%
\providecommand \@@endlink[0]{}%
\providecommand \url  [0]{\begingroup\@sanitize@url \@url }%
\providecommand \@url [1]{\endgroup\@href {#1}{\urlprefix }}%
\providecommand \urlprefix  [0]{URL }%
\providecommand \Eprint [0]{\href }%
\providecommand \doibase [0]{http://dx.doi.org/}%
\providecommand \selectlanguage [0]{\@gobble}%
\providecommand \bibinfo  [0]{\@secondoftwo}%
\providecommand \bibfield  [0]{\@secondoftwo}%
\providecommand \translation [1]{[#1]}%
\providecommand \BibitemOpen [0]{}%
\providecommand \bibitemStop [0]{}%
\providecommand \bibitemNoStop [0]{.\EOS\space}%
\providecommand \EOS [0]{\spacefactor3000\relax}%
\providecommand \BibitemShut  [1]{\csname bibitem#1\endcsname}%
\let\auto@bib@innerbib\@empty
\bibitem [{\citenamefont {Strominger}\ and\ \citenamefont
  {Vafa}(1996)}]{Strominger:1996sh}%
  \BibitemOpen
  \bibfield  {author} {\bibinfo {author} {\bibfnamefont {A.}~\bibnamefont
  {Strominger}}\ and\ \bibinfo {author} {\bibfnamefont {C.}~\bibnamefont
  {Vafa}},\ }\href {\doibase 10.1016/0370-2693(96)00345-0} {\bibfield
  {journal} {\bibinfo  {journal} {Phys. Lett. B}\ }\textbf {\bibinfo {volume}
  {379}},\ \bibinfo {pages} {99} (\bibinfo {year} {1996})},\ \Eprint
  {http://arxiv.org/abs/hep-th/9601029} {arXiv:hep-th/9601029} \BibitemShut
  {NoStop}%
\bibitem [{\citenamefont {Benini}\ \emph {et~al.}(2016)\citenamefont {Benini},
  \citenamefont {Hristov},\ and\ \citenamefont {Zaffaroni}}]{Benini:2015eyy}%
  \BibitemOpen
  \bibfield  {author} {\bibinfo {author} {\bibfnamefont {F.}~\bibnamefont
  {Benini}}, \bibinfo {author} {\bibfnamefont {K.}~\bibnamefont {Hristov}}, \
  and\ \bibinfo {author} {\bibfnamefont {A.}~\bibnamefont {Zaffaroni}},\ }\href
  {\doibase 10.1007/JHEP05(2016)054} {\bibfield  {journal} {\bibinfo  {journal}
  {JHEP}\ }\textbf {\bibinfo {volume} {05}},\ \bibinfo {pages} {054} (\bibinfo
  {year} {2016})},\ \Eprint {http://arxiv.org/abs/1511.04085} {arXiv:1511.04085
  [hep-th]} \BibitemShut {NoStop}%
\bibitem [{\citenamefont {Benini}\ \emph {et~al.}(2017)\citenamefont {Benini},
  \citenamefont {Hristov},\ and\ \citenamefont {Zaffaroni}}]{Benini:2016rke}%
  \BibitemOpen
  \bibfield  {author} {\bibinfo {author} {\bibfnamefont {F.}~\bibnamefont
  {Benini}}, \bibinfo {author} {\bibfnamefont {K.}~\bibnamefont {Hristov}}, \
  and\ \bibinfo {author} {\bibfnamefont {A.}~\bibnamefont {Zaffaroni}},\ }\href
  {\doibase 10.1016/j.physletb.2017.05.076} {\bibfield  {journal} {\bibinfo
  {journal} {Phys. Lett.}\ }\textbf {\bibinfo {volume} {B771}},\ \bibinfo
  {pages} {462} (\bibinfo {year} {2017})},\ \Eprint
  {http://arxiv.org/abs/1608.07294} {arXiv:1608.07294 [hep-th]} \BibitemShut
  {NoStop}%
\bibitem [{\citenamefont {Couzens}\ \emph {et~al.}(2019)\citenamefont
  {Couzens}, \citenamefont {Gauntlett}, \citenamefont {Martelli},\ and\
  \citenamefont {Sparks}}]{Couzens:2018wnk}%
  \BibitemOpen
  \bibfield  {author} {\bibinfo {author} {\bibfnamefont {C.}~\bibnamefont
  {Couzens}}, \bibinfo {author} {\bibfnamefont {J.~P.}\ \bibnamefont
  {Gauntlett}}, \bibinfo {author} {\bibfnamefont {D.}~\bibnamefont {Martelli}},
  \ and\ \bibinfo {author} {\bibfnamefont {J.}~\bibnamefont {Sparks}},\ }\href
  {\doibase 10.1007/JHEP01(2019)212} {\bibfield  {journal} {\bibinfo  {journal}
  {JHEP}\ }\textbf {\bibinfo {volume} {01}},\ \bibinfo {pages} {212} (\bibinfo
  {year} {2019})},\ \Eprint {http://arxiv.org/abs/1810.11026} {arXiv:1810.11026
  [hep-th]} \BibitemShut {NoStop}%
\bibitem [{\citenamefont {Gauntlett}\ \emph {et~al.}(2019)\citenamefont
  {Gauntlett}, \citenamefont {Martelli},\ and\ \citenamefont
  {Sparks}}]{Gauntlett:2019roi}%
  \BibitemOpen
  \bibfield  {author} {\bibinfo {author} {\bibfnamefont {J.~P.}\ \bibnamefont
  {Gauntlett}}, \bibinfo {author} {\bibfnamefont {D.}~\bibnamefont {Martelli}},
  \ and\ \bibinfo {author} {\bibfnamefont {J.}~\bibnamefont {Sparks}},\ }\href
  {\doibase 10.1007/JHEP06(2019)140} {\bibfield  {journal} {\bibinfo  {journal}
  {JHEP}\ }\textbf {\bibinfo {volume} {06}},\ \bibinfo {pages} {140} (\bibinfo
  {year} {2019})},\ \Eprint {http://arxiv.org/abs/1904.04282} {arXiv:1904.04282
  [hep-th]} \BibitemShut {NoStop}%
\bibitem [{\citenamefont {Hosseini}\ and\ \citenamefont
  {Zaffaroni}(2019{\natexlab{a}})}]{Hosseini:2019ddy}%
  \BibitemOpen
  \bibfield  {author} {\bibinfo {author} {\bibfnamefont {S.~M.}\ \bibnamefont
  {Hosseini}}\ and\ \bibinfo {author} {\bibfnamefont {A.}~\bibnamefont
  {Zaffaroni}},\ }\href {\doibase 10.1007/JHEP07(2019)174} {\bibfield
  {journal} {\bibinfo  {journal} {JHEP}\ }\textbf {\bibinfo {volume} {07}},\
  \bibinfo {pages} {174} (\bibinfo {year} {2019}{\natexlab{a}})},\ \Eprint
  {http://arxiv.org/abs/1904.04269} {arXiv:1904.04269 [hep-th]} \BibitemShut
  {NoStop}%
\bibitem [{\citenamefont {Kim}\ and\ \citenamefont {Kim}(2019)}]{Kim:2019umc}%
  \BibitemOpen
  \bibfield  {author} {\bibinfo {author} {\bibfnamefont {H.}~\bibnamefont
  {Kim}}\ and\ \bibinfo {author} {\bibfnamefont {N.}~\bibnamefont {Kim}},\
  }\href {\doibase 10.1007/JHEP11(2019)050} {\bibfield  {journal} {\bibinfo
  {journal} {JHEP}\ }\textbf {\bibinfo {volume} {11}},\ \bibinfo {pages} {050}
  (\bibinfo {year} {2019})},\ \Eprint {http://arxiv.org/abs/1904.05344}
  {arXiv:1904.05344 [hep-th]} \BibitemShut {NoStop}%
\bibitem [{\citenamefont {Sen}(2005)}]{Sen:2005wa}%
  \BibitemOpen
  \bibfield  {author} {\bibinfo {author} {\bibfnamefont {A.}~\bibnamefont
  {Sen}},\ }\href {\doibase 10.1088/1126-6708/2005/09/038} {\bibfield
  {journal} {\bibinfo  {journal} {JHEP}\ }\textbf {\bibinfo {volume} {09}},\
  \bibinfo {pages} {038} (\bibinfo {year} {2005})},\ \Eprint
  {http://arxiv.org/abs/hep-th/0506177} {arXiv:hep-th/0506177} \BibitemShut
  {NoStop}%
\bibitem [{\citenamefont {Hosseini}\ and\ \citenamefont
  {Zaffaroni}(2016)}]{Hosseini:2016tor}%
  \BibitemOpen
  \bibfield  {author} {\bibinfo {author} {\bibfnamefont {S.~M.}\ \bibnamefont
  {Hosseini}}\ and\ \bibinfo {author} {\bibfnamefont {A.}~\bibnamefont
  {Zaffaroni}},\ }\href {\doibase 10.1007/JHEP08(2016)064} {\bibfield
  {journal} {\bibinfo  {journal} {JHEP}\ }\textbf {\bibinfo {volume} {08}},\
  \bibinfo {pages} {064} (\bibinfo {year} {2016})},\ \Eprint
  {http://arxiv.org/abs/1604.03122} {arXiv:1604.03122 [hep-th]} \BibitemShut
  {NoStop}%
\bibitem [{\citenamefont {Hosseini}\ and\ \citenamefont
  {Mekareeya}(2016)}]{Hosseini:2016ume}%
  \BibitemOpen
  \bibfield  {author} {\bibinfo {author} {\bibfnamefont {S.~M.}\ \bibnamefont
  {Hosseini}}\ and\ \bibinfo {author} {\bibfnamefont {N.}~\bibnamefont
  {Mekareeya}},\ }\href {\doibase 10.1007/JHEP08(2016)089} {\bibfield
  {journal} {\bibinfo  {journal} {JHEP}\ }\textbf {\bibinfo {volume} {08}},\
  \bibinfo {pages} {089} (\bibinfo {year} {2016})},\ \Eprint
  {http://arxiv.org/abs/1604.03397} {arXiv:1604.03397 [hep-th]} \BibitemShut
  {NoStop}%
\bibitem [{\citenamefont {Ferrero}\ \emph {et~al.}(2021)\citenamefont
  {Ferrero}, \citenamefont {Gauntlett}, \citenamefont {Ipi\~na}, \citenamefont
  {Martelli},\ and\ \citenamefont {Sparks}}]{Ferrero:2020twa}%
  \BibitemOpen
  \bibfield  {author} {\bibinfo {author} {\bibfnamefont {P.}~\bibnamefont
  {Ferrero}}, \bibinfo {author} {\bibfnamefont {J.~P.}\ \bibnamefont
  {Gauntlett}}, \bibinfo {author} {\bibfnamefont {J.~M.~P.}\ \bibnamefont
  {Ipi\~na}}, \bibinfo {author} {\bibfnamefont {D.}~\bibnamefont {Martelli}}, \
  and\ \bibinfo {author} {\bibfnamefont {J.}~\bibnamefont {Sparks}},\ }\href
  {\doibase 10.1103/PhysRevD.104.046007} {\bibfield  {journal} {\bibinfo
  {journal} {Phys. Rev. D}\ }\textbf {\bibinfo {volume} {104}},\ \bibinfo
  {pages} {046007} (\bibinfo {year} {2021})},\ \Eprint
  {http://arxiv.org/abs/2012.08530} {arXiv:2012.08530 [hep-th]} \BibitemShut
  {NoStop}%
\bibitem [{\citenamefont {Hosseini}\ \emph {et~al.}(2019)\citenamefont
  {Hosseini}, \citenamefont {Hristov},\ and\ \citenamefont
  {Zaffaroni}}]{Hosseini:2019iad}%
  \BibitemOpen
  \bibfield  {author} {\bibinfo {author} {\bibfnamefont {S.~M.}\ \bibnamefont
  {Hosseini}}, \bibinfo {author} {\bibfnamefont {K.}~\bibnamefont {Hristov}}, \
  and\ \bibinfo {author} {\bibfnamefont {A.}~\bibnamefont {Zaffaroni}},\ }\href
  {\doibase 10.1007/JHEP12(2019)168} {\bibfield  {journal} {\bibinfo  {journal}
  {JHEP}\ }\textbf {\bibinfo {volume} {12}},\ \bibinfo {pages} {168} (\bibinfo
  {year} {2019})},\ \Eprint {http://arxiv.org/abs/1909.10550} {arXiv:1909.10550
  [hep-th]} \BibitemShut {NoStop}%
\bibitem [{\citenamefont {Hosseini}\ \emph {et~al.}(2021)\citenamefont
  {Hosseini}, \citenamefont {Hristov},\ and\ \citenamefont
  {Zaffaroni}}]{Hosseini:2021fge}%
  \BibitemOpen
  \bibfield  {author} {\bibinfo {author} {\bibfnamefont {S.~M.}\ \bibnamefont
  {Hosseini}}, \bibinfo {author} {\bibfnamefont {K.}~\bibnamefont {Hristov}}, \
  and\ \bibinfo {author} {\bibfnamefont {A.}~\bibnamefont {Zaffaroni}},\ }\href
  {\doibase 10.1007/JHEP07(2021)182} {\bibfield  {journal} {\bibinfo  {journal}
  {JHEP}\ }\textbf {\bibinfo {volume} {07}},\ \bibinfo {pages} {182} (\bibinfo
  {year} {2021})},\ \Eprint {http://arxiv.org/abs/2104.11249} {arXiv:2104.11249
  [hep-th]} \BibitemShut {NoStop}%
\bibitem [{\citenamefont {Faedo}\ and\ \citenamefont
  {Martelli}(2022)}]{Faedo:2021nub}%
  \BibitemOpen
  \bibfield  {author} {\bibinfo {author} {\bibfnamefont {F.}~\bibnamefont
  {Faedo}}\ and\ \bibinfo {author} {\bibfnamefont {D.}~\bibnamefont
  {Martelli}},\ }\href {\doibase 10.1007/JHEP02(2022)101} {\bibfield  {journal}
  {\bibinfo  {journal} {JHEP}\ }\textbf {\bibinfo {volume} {02}},\ \bibinfo
  {pages} {101} (\bibinfo {year} {2022})},\ \Eprint
  {http://arxiv.org/abs/2111.13660} {arXiv:2111.13660 [hep-th]} \BibitemShut
  {NoStop}%
\bibitem [{Note1()}]{Note1}%
  \BibitemOpen
  \bibinfo {note} {It would be of much interest to generalise the entropy
  functions of this paper to incorporate higher-derivative corrections and
  hence go beyond the large $N$ limit.}\BibitemShut {Stop}%
\bibitem [{\citenamefont {Kim}\ and\ \citenamefont {Park}(2006)}]{Kim:2006qu}%
  \BibitemOpen
  \bibfield  {author} {\bibinfo {author} {\bibfnamefont {N.}~\bibnamefont
  {Kim}}\ and\ \bibinfo {author} {\bibfnamefont {J.-D.}\ \bibnamefont {Park}},\
  }\href {\doibase 10.1088/1126-6708/2006/09/041} {\bibfield  {journal}
  {\bibinfo  {journal} {JHEP}\ }\textbf {\bibinfo {volume} {09}},\ \bibinfo
  {pages} {041} (\bibinfo {year} {2006})},\ \Eprint
  {http://arxiv.org/abs/hep-th/0607093} {arXiv:hep-th/0607093} \BibitemShut
  {NoStop}%
\bibitem [{\citenamefont {Gauntlett}\ and\ \citenamefont
  {Kim}(2008)}]{Gauntlett:2007ts}%
  \BibitemOpen
  \bibfield  {author} {\bibinfo {author} {\bibfnamefont {J.~P.}\ \bibnamefont
  {Gauntlett}}\ and\ \bibinfo {author} {\bibfnamefont {N.}~\bibnamefont
  {Kim}},\ }\href {\doibase 10.1007/s00220-008-0575-5} {\bibfield  {journal}
  {\bibinfo  {journal} {Commun. Math. Phys.}\ }\textbf {\bibinfo {volume}
  {284}},\ \bibinfo {pages} {897} (\bibinfo {year} {2008})},\ \Eprint
  {http://arxiv.org/abs/0710.2590} {arXiv:0710.2590 [hep-th]} \BibitemShut
  {NoStop}%
\bibitem [{\citenamefont {Benvenuti}\ \emph {et~al.}(2006)\citenamefont
  {Benvenuti}, \citenamefont {Pando~Zayas},\ and\ \citenamefont
  {Tachikawa}}]{Benvenuti:2006xg}%
  \BibitemOpen
  \bibfield  {author} {\bibinfo {author} {\bibfnamefont {S.}~\bibnamefont
  {Benvenuti}}, \bibinfo {author} {\bibfnamefont {L.~A.}\ \bibnamefont
  {Pando~Zayas}}, \ and\ \bibinfo {author} {\bibfnamefont {Y.}~\bibnamefont
  {Tachikawa}},\ }\href {\doibase 10.4310/ATMP.2006.v10.n3.a4} {\bibfield
  {journal} {\bibinfo  {journal} {Adv. Theor. Math. Phys.}\ }\textbf {\bibinfo
  {volume} {10}},\ \bibinfo {pages} {395} (\bibinfo {year} {2006})},\ \Eprint
  {http://arxiv.org/abs/hep-th/0601054} {arXiv:hep-th/0601054} \BibitemShut
  {NoStop}%
\bibitem [{Note2()}]{Note2}%
  \BibitemOpen
  \bibinfo {note} {While this is generically expected to be true, it is
  important to understand the necessary and sufficient conditions.}\BibitemShut
  {Stop}%
\bibitem [{Note3()}]{Note3}%
  \BibitemOpen
  \bibinfo {note} {$Y_9$ is free of orbifold singularities provided the $p_i$
  are coprime to both of $m_\pm $.}\BibitemShut {Stop}%
\bibitem [{\citenamefont {Ferrero}\ \emph
  {et~al.}(2022{\natexlab{a}})\citenamefont {Ferrero}, \citenamefont
  {Gauntlett},\ and\ \citenamefont {Sparks}}]{Ferrero:2021etw}%
  \BibitemOpen
  \bibfield  {author} {\bibinfo {author} {\bibfnamefont {P.}~\bibnamefont
  {Ferrero}}, \bibinfo {author} {\bibfnamefont {J.~P.}\ \bibnamefont
  {Gauntlett}}, \ and\ \bibinfo {author} {\bibfnamefont {J.}~\bibnamefont
  {Sparks}},\ }\href {\doibase 10.1007/JHEP01(2022)102} {\bibfield  {journal}
  {\bibinfo  {journal} {JHEP}\ }\textbf {\bibinfo {volume} {01}},\ \bibinfo
  {pages} {102} (\bibinfo {year} {2022}{\natexlab{a}})},\ \Eprint
  {http://arxiv.org/abs/2112.01543} {arXiv:2112.01543 [hep-th]} \BibitemShut
  {NoStop}%
\bibitem [{Note4()}]{Note4}%
  \BibitemOpen
  \bibinfo {note} {We must have $N=m_+N^{X_+}=m_-N^{X_-}$ with $N^{X_\pm }\in
  \protect \mathbb {Z}$ and hence $N=m_+ m_-{\protect \mathcal N}_0$ with
  ${\protect \mathcal N}_0\in \protect \mathbb {Z}$ \cite
  {Boido:2022mbe}.}\BibitemShut {Stop}%
\bibitem [{Note5()}]{Note5}%
  \BibitemOpen
  \bibinfo {note} {The normalization factor of $N$ is due to the fact that
  baryonic operators, dual to M5-branes wrapped on $\Sigma _I$, arise as
  $N\times N$ determinants in Chern-Simons-matter duals, and $P_I$ is then the
  charge of the associated field.}\BibitemShut {Stop}%
\bibitem [{\citenamefont {Boido}\ \emph {et~al.}(2022)\citenamefont {Boido},
  \citenamefont {Gauntlett}, \citenamefont {Martelli},\ and\ \citenamefont
  {Sparks}}]{Boido:2022mbe}%
  \BibitemOpen
  \bibfield  {author} {\bibinfo {author} {\bibfnamefont {A.}~\bibnamefont
  {Boido}}, \bibinfo {author} {\bibfnamefont {J.~P.}\ \bibnamefont
  {Gauntlett}}, \bibinfo {author} {\bibfnamefont {D.}~\bibnamefont {Martelli}},
  \ and\ \bibinfo {author} {\bibfnamefont {J.}~\bibnamefont {Sparks}},\
  }\href@noop {} {\  (\bibinfo {year} {2022})},\ \Eprint
  {http://arxiv.org/abs/2211.02662} {arXiv:2211.02662 [hep-th]} \BibitemShut
  {NoStop}%
\bibitem [{Note6()}]{Note6}%
  \BibitemOpen
  \bibinfo {note} {More precisely these fibres are $X_7/\protect \mathbb
  {Z}_{m_\pm }$, so $X_7^\pm $ are really covering spaces of the fibres. The
  orientations of $X_7^\pm $ are discussed in more detail in \cite
  {Boido:2022mbe}.}\BibitemShut {Stop}%
\bibitem [{Note7()}]{Note7}%
  \BibitemOpen
  \bibinfo {note} {Tools for analysing toric examples are developed in \cite
  {Boido:2022mbe}}\BibitemShut {NoStop}%
\bibitem [{\citenamefont {Martelli}\ \emph {et~al.}(2008)\citenamefont
  {Martelli}, \citenamefont {Sparks},\ and\ \citenamefont
  {Yau}}]{Martelli:2006yb}%
  \BibitemOpen
  \bibfield  {author} {\bibinfo {author} {\bibfnamefont {D.}~\bibnamefont
  {Martelli}}, \bibinfo {author} {\bibfnamefont {J.}~\bibnamefont {Sparks}}, \
  and\ \bibinfo {author} {\bibfnamefont {S.-T.}\ \bibnamefont {Yau}},\ }\href
  {\doibase 10.1007/s00220-008-0479-4} {\bibfield  {journal} {\bibinfo
  {journal} {Commun. Math. Phys.}\ }\textbf {\bibinfo {volume} {280}},\
  \bibinfo {pages} {611} (\bibinfo {year} {2008})},\ \Eprint
  {http://arxiv.org/abs/hep-th/0603021} {arXiv:hep-th/0603021 [hep-th]}
  \BibitemShut {NoStop}%
\bibitem [{\citenamefont {Martelli}\ and\ \citenamefont
  {Sparks}(2011)}]{Martelli:2011qj}%
  \BibitemOpen
  \bibfield  {author} {\bibinfo {author} {\bibfnamefont {D.}~\bibnamefont
  {Martelli}}\ and\ \bibinfo {author} {\bibfnamefont {J.}~\bibnamefont
  {Sparks}},\ }\href {\doibase 10.1103/PhysRevD.84.046008} {\bibfield
  {journal} {\bibinfo  {journal} {Phys. Rev.}\ }\textbf {\bibinfo {volume}
  {D84}},\ \bibinfo {pages} {046008} (\bibinfo {year} {2011})},\ \Eprint
  {http://arxiv.org/abs/1102.5289} {arXiv:1102.5289 [hep-th]} \BibitemShut
  {NoStop}%
\bibitem [{\citenamefont {Cheon}\ \emph {et~al.}(2011)\citenamefont {Cheon},
  \citenamefont {Kim},\ and\ \citenamefont {Kim}}]{Cheon:2011vi}%
  \BibitemOpen
  \bibfield  {author} {\bibinfo {author} {\bibfnamefont {S.}~\bibnamefont
  {Cheon}}, \bibinfo {author} {\bibfnamefont {H.}~\bibnamefont {Kim}}, \ and\
  \bibinfo {author} {\bibfnamefont {N.}~\bibnamefont {Kim}},\ }\href {\doibase
  10.1007/JHEP05(2011)134} {\bibfield  {journal} {\bibinfo  {journal} {JHEP}\
  }\textbf {\bibinfo {volume} {05}},\ \bibinfo {pages} {134} (\bibinfo {year}
  {2011})},\ \Eprint {http://arxiv.org/abs/1102.5565} {arXiv:1102.5565
  [hep-th]} \BibitemShut {NoStop}%
\bibitem [{\citenamefont {Jafferis}\ \emph {et~al.}(2011)\citenamefont
  {Jafferis}, \citenamefont {Klebanov}, \citenamefont {Pufu},\ and\
  \citenamefont {Safdi}}]{Jafferis:2011zi}%
  \BibitemOpen
  \bibfield  {author} {\bibinfo {author} {\bibfnamefont {D.~L.}\ \bibnamefont
  {Jafferis}}, \bibinfo {author} {\bibfnamefont {I.~R.}\ \bibnamefont
  {Klebanov}}, \bibinfo {author} {\bibfnamefont {S.~S.}\ \bibnamefont {Pufu}},
  \ and\ \bibinfo {author} {\bibfnamefont {B.~R.}\ \bibnamefont {Safdi}},\
  }\href {\doibase 10.1007/JHEP06(2011)102} {\bibfield  {journal} {\bibinfo
  {journal} {JHEP}\ }\textbf {\bibinfo {volume} {06}},\ \bibinfo {pages} {102}
  (\bibinfo {year} {2011})},\ \Eprint {http://arxiv.org/abs/1103.1181}
  {arXiv:1103.1181 [hep-th]} \BibitemShut {NoStop}%
\bibitem [{\citenamefont {Couzens}(2022)}]{Couzens:2021cpk}%
  \BibitemOpen
  \bibfield  {author} {\bibinfo {author} {\bibfnamefont {C.}~\bibnamefont
  {Couzens}},\ }\href {\doibase 10.1007/JHEP03(2022)078} {\bibfield  {journal}
  {\bibinfo  {journal} {JHEP}\ }\textbf {\bibinfo {volume} {03}},\ \bibinfo
  {pages} {078} (\bibinfo {year} {2022})},\ \Eprint
  {http://arxiv.org/abs/2112.04462} {arXiv:2112.04462 [hep-th]} \BibitemShut
  {NoStop}%
\bibitem [{\citenamefont {Hosseini}\ and\ \citenamefont
  {Zaffaroni}(2019{\natexlab{b}})}]{Hosseini:2019use}%
  \BibitemOpen
  \bibfield  {author} {\bibinfo {author} {\bibfnamefont {S.~M.}\ \bibnamefont
  {Hosseini}}\ and\ \bibinfo {author} {\bibfnamefont {A.}~\bibnamefont
  {Zaffaroni}},\ }\href {\doibase 10.1007/JHEP03(2019)108} {\bibfield
  {journal} {\bibinfo  {journal} {JHEP}\ }\textbf {\bibinfo {volume} {03}},\
  \bibinfo {pages} {108} (\bibinfo {year} {2019}{\natexlab{b}})},\ \Eprint
  {http://arxiv.org/abs/1901.05977} {arXiv:1901.05977 [hep-th]} \BibitemShut
  {NoStop}%
\bibitem [{Note8()}]{Note8}%
  \BibitemOpen
  \bibinfo {note} {Switching off rotation and electric charge gives solutions
  where the spindle degenerates at the $AdS_4$ boundary \cite
  {Ferrero:2020twa}.}\BibitemShut {Stop}%
\bibitem [{\citenamefont {Cassani}\ \emph {et~al.}(2021)\citenamefont
  {Cassani}, \citenamefont {Gauntlett}, \citenamefont {Martelli},\ and\
  \citenamefont {Sparks}}]{Cassani:2021dwa}%
  \BibitemOpen
  \bibfield  {author} {\bibinfo {author} {\bibfnamefont {D.}~\bibnamefont
  {Cassani}}, \bibinfo {author} {\bibfnamefont {J.~P.}\ \bibnamefont
  {Gauntlett}}, \bibinfo {author} {\bibfnamefont {D.}~\bibnamefont {Martelli}},
  \ and\ \bibinfo {author} {\bibfnamefont {J.}~\bibnamefont {Sparks}},\ }\href
  {\doibase 10.1103/PhysRevD.104.086005} {\bibfield  {journal} {\bibinfo
  {journal} {Phys. Rev. D}\ }\textbf {\bibinfo {volume} {104}},\ \bibinfo
  {pages} {086005} (\bibinfo {year} {2021})},\ \Eprint
  {http://arxiv.org/abs/2106.05571} {arXiv:2106.05571 [hep-th]} \BibitemShut
  {NoStop}%
\bibitem [{\citenamefont {Cassani}\ and\ \citenamefont
  {Papini}(2019)}]{Cassani:2019mms}%
  \BibitemOpen
  \bibfield  {author} {\bibinfo {author} {\bibfnamefont {D.}~\bibnamefont
  {Cassani}}\ and\ \bibinfo {author} {\bibfnamefont {L.}~\bibnamefont
  {Papini}},\ }\href {\doibase 10.1007/JHEP09(2019)079} {\bibfield  {journal}
  {\bibinfo  {journal} {JHEP}\ }\textbf {\bibinfo {volume} {09}},\ \bibinfo
  {pages} {079} (\bibinfo {year} {2019})},\ \Eprint
  {http://arxiv.org/abs/1906.10148} {arXiv:1906.10148 [hep-th]} \BibitemShut
  {NoStop}%
\bibitem [{\citenamefont {Ferrero}\ \emph
  {et~al.}(2022{\natexlab{b}})\citenamefont {Ferrero}, \citenamefont {Inglese},
  \citenamefont {Martelli},\ and\ \citenamefont {Sparks}}]{Ferrero:2021ovq}%
  \BibitemOpen
  \bibfield  {author} {\bibinfo {author} {\bibfnamefont {P.}~\bibnamefont
  {Ferrero}}, \bibinfo {author} {\bibfnamefont {M.}~\bibnamefont {Inglese}},
  \bibinfo {author} {\bibfnamefont {D.}~\bibnamefont {Martelli}}, \ and\
  \bibinfo {author} {\bibfnamefont {J.}~\bibnamefont {Sparks}},\ }\href
  {\doibase 10.1103/PhysRevD.105.126001} {\bibfield  {journal} {\bibinfo
  {journal} {Phys. Rev. D}\ }\textbf {\bibinfo {volume} {105}},\ \bibinfo
  {pages} {126001} (\bibinfo {year} {2022}{\natexlab{b}})},\ \Eprint
  {http://arxiv.org/abs/2109.14625} {arXiv:2109.14625 [hep-th]} \BibitemShut
  {NoStop}%
\bibitem [{\citenamefont {Couzens}\ \emph {et~al.}(2021)\citenamefont
  {Couzens}, \citenamefont {Marcus}, \citenamefont {Stemerdink},\ and\
  \citenamefont {van~de Heisteeg}}]{Couzens:2020jgx}%
  \BibitemOpen
  \bibfield  {author} {\bibinfo {author} {\bibfnamefont {C.}~\bibnamefont
  {Couzens}}, \bibinfo {author} {\bibfnamefont {E.}~\bibnamefont {Marcus}},
  \bibinfo {author} {\bibfnamefont {K.}~\bibnamefont {Stemerdink}}, \ and\
  \bibinfo {author} {\bibfnamefont {D.}~\bibnamefont {van~de Heisteeg}},\
  }\href {\doibase 10.1007/JHEP05(2021)194} {\bibfield  {journal} {\bibinfo
  {journal} {JHEP}\ }\textbf {\bibinfo {volume} {05}},\ \bibinfo {pages} {194}
  (\bibinfo {year} {2021})},\ \Eprint {http://arxiv.org/abs/2011.07071}
  {arXiv:2011.07071 [hep-th]} \BibitemShut {NoStop}%
\bibitem [{Note9()}]{Note9}%
  \BibitemOpen
  \bibinfo {note} {\cite {Boido:2022mbe} also shows that associated with
  $U(1)^s$ invariant non-trivial five-cycles $\Sigma _I$ on $X_7$, (16) can be
  recast in the form $P_I = \protect \frac {1}{2b_0}\left (R_I^+-R_I^-\right
  )$}\BibitemShut {NoStop}%
\end{thebibliography}%

\end{document}